\documentclass[final,1p,times]{elsarticle}
\usepackage{graphics}
\usepackage{graphicx}
\usepackage{amssymb}
\usepackage{amsmath} %
\biboptions{sort&compress}
\usepackage{lineno}
\usepackage{lscape}
\usepackage{color}
\journal{Nuclear Instruments and Methods A}

\begin{document}
%\begin{linenumbers}
\begin{frontmatter}



\title{Measurement of the radon concentration in purified water in the Super-Kamiokande IV detector}

% if there is only one institution, use this form:
%\author{John Author, Giovanna Autore}
%\address{University of Wisdom, Physics City, Scienceland}

% else, use optional labels to link authors explicitly to addresses,
% as shown below:
\author[A]{Y.~Nakano\corref{cor1}}
\ead{ynakano@phys.sci.kobe-u.ac.jp}
\author[B]{T.~Hokama}
\author[C]{M.~Matsubara}
\author[D]{M.~Miwa}
\author[E,F]{M.~Nakahata}
\author[G]{T.~Nakamura}
\author[E,F]{H.~Sekiya}
\author[A,F]{Y.~Takeuchi}
\author[E]{S.~Tasaka}
\author[H,F]{R.A.~Wendell}

\address[A]{Department of Physics, Graduate School of Science, Kobe University, Kobe, Hyogo 657-8501, Japan}
\address[B]{Nuclear Emergency Assistance and Training Center, Japan Atomic Energy Agency, Ibaraki 311-1206, Japan}
\address[C]{Head Office for Information and Management, Gifu University, Gifu 501-1193, Japan}
\address[D]{Division of Radioisotope Experiment, Life Science Research Center, Gifu University, Gifu 501-1193, Japan}
\address[E]{Kamioka Observatory, Institute for Cosmic Ray Research, The University of Tokyo, Gifu 506-1205, Japan}
\address[F]{Kavli Institute for the Physics and Mathematics of the Universe (WPI), \\ The University of
Tokyo Institutes for Advanced Study, The University of Tokyo, Kashiwa, Chiba 277-8583, Japan}
\address[G]{Department of Physics, Faculty of Education, Gifu University, Gifu 501-1193, Japan}
\address[H]{Department of Physics, Kyoto University, Kyoto, Kyoto 606-8502, Japan}

\cortext[cor1]{Corresponding author. Tel: +81 78 803 5640; Fax: +81 78 803 5662.}

\begin{abstract}

The radioactive noble gas radon can be a serious background source in the underground particle physics experiments studying processes that deposit energy comparable to its decay products. 
Low energy solar neutrino measurements at Super-Kamiokande suffer from these backgrounds and therefore require precise characterization of the radon concentration in the detector's ultra-pure water.
For this purpose, we have developed a measurement system consisting of a radon extraction column, a charcoal trap, and a radon detector. 
In this article we discuss the design, calibration, and performance of the radon extraction column. 
We also describe the design of the measurement system and evaluate its performance, including its background. 
Using this system we measured the radon concentration in Super-Kamiokande's water between May 2014 and October 2015. 
The measured radon concentration in the supply lines of the water circulation system was $1.74\pm0.14~\mathrm{mBq/m^{3}}$ and in the return line was $9.06\pm0.58~\mathrm{mBq/m^{3}}$. Water sampled from the center region of the detector itself had a concentration of $<0.23~\mathrm{mBq/m^{3}}$~($95\%$~C.L.) and water sampled from the bottom region of the detector had a concentration of $2.63\pm0.22~\mathrm{mBq/m^{3}}$.
\end{abstract}

\begin{keyword}
Super-Kamiokande \sep Radon \sep Solar neutrino \sep Charcoal \sep Radon extraction column 
%\sep
\end{keyword}

\end{frontmatter}

%\linenumbers

\section{Introduction}
The framework of three-flavor neutrino oscillations~\cite{osci1, osci2} is increasingly well understood.  However, there remain unknown quantities, including the absolute scale of the neutrino mass states and whether CP symmetry is violated in neutrino mixing.
Hints of oscillations among solar neutrinos were first obtained from the difference between the solar-neutrino fluxes measured with the elastic-scattering channel at Super-Kamiokande~(Super-K) 
and the charged-current channel at the Sudbury Neutrino Observatory~(SNO) in 2001~\cite{sk1_es, sno1}.
 Though only electron neutrinos are produced in the core of the Sun, this result demonstrated the existence of other neutrino components in the solar neutrino flux.
 Solar neutrino oscillations were subsequently confirmed by including neutral-current measurements from SNO~\cite{sno2}.
 Precise solar neutrino oscillation measurements require large statistics and Super-K  provided such measurements with its large detector volume~\cite{sk1,sk2,sk3,sk4_solar}.
 Its large size further benefits Super-K's efforts to test for the presence of solar and terrestrial matter effects in the solar neutrino oscillations as predicted by Mikheyev, Smirnov and Wolfenstein~\cite{msw1,msw2}, the so-called MSW effect.
 Although the MSW effect is enough to explain the current solar neutrino data, direct evidence for it has not yet been obtained.
 At Super-K, it manifests as a distortion in the energy spectrum of recoiling electrons produced by solar neutrino interactions in water.
 Measurements of the electron recoil energy spectrum and the asymmetry of the solar neutrino interaction rate  during the day and during the night~\cite{dn_sk4} 
allow Super-K to directly probe the matter effects in the Sun and Earth, respectively.

Moreover, this distortion is expected to increase at lower energies~(the so-called \lq\lq upturn\rq\rq) and therefore requires an energy threshold that is as low as possible.
At Super-K, the energy threshold is limited by radioactive backgrounds, most of which arise from radon~(Rn) contamination in the detector water~\cite{Rn_sk}. 
 In order to achieve a suitably low background for solar neutrino analysis at the lowest energies, the Rn emanation from all components of the Super-K detector must be reduced.
 For this purpose, we have developed a new measurement system to monitor the Rn concentration in purified water.
Herein we present details of the system as well as results of in-situ measurements at Super-K.

This paper is organized as follows.
 In Section~\ref{sk_det} we describe the Super-K detector and its water system as well as summarize the history of Rn studies at Super-K.
 Section~\ref{sec_2}  presents a new method of extracting Rn from water and subsequently describes the design and data analysis of a system to measure the extracted Rn. 
 We discuss background levels and systematic uncertainties accompanying this measurement in Section~\ref{sec_bg} before 
 presenting measurement results in Section~\ref{sec_measure}.
 In Section~\ref{sec_5}, we discuss several possible Rn sources in the Super-K tank.
 Finally in Section~\ref{sec_6}, we conclude this study and outline prospects for the future.
 Note that we use the term Rn to refer specifically to $\mathrm{^{222}Rn}$ in this paper, unless otherwise stated.

\section{The Super-Kamiokande detector} \label{sk_det}

Super-K is a water Cherenkov detector containing $50000$~tonnes of highly purified water that is viewed by 11129 20-inch photomultiplier tubes~(PMTs)~\cite{paper_SK}.
 It is located roughly $1000$~m underground ($2700$~m water equivalent) inside the Ikenoyama mountain in Gifu prefecture, Japan.
 We define the local coordinate system of the detector as $(x,y,z)$, where $(x,y)$ represents the plane of the cylinder as viewed from above and $z$ represents the height within the detector tank~\cite{paper_SK}.
The origin is placed at the center of the tank.

After the installation of new front end electronics in 2008~\cite{yamada}, the fourth phase of Super-K~(SK-IV) started and was concluded 10 years later in May 2018.
 With improvements in the water circulation system, calibration methods~\cite{Calib_p}, and event selection, 
the detector's energy threshold was lowered to $3.5$~MeV in terms of electron recoil kinetic energy~\cite{sk4_solar}.
 These upgrades have also allowed for precision measurements of the electron recoil energy spectrum  
and the Super-K data together with data from SNO provide the strongest constraint on the solar electron neutrino survival probability~\cite{sk4_solar,sno}. 
 However, experimental data from other solar neutrino experiments, including radiochemical experiments~\cite{home,sage,gno}, water Cherenkov experiments~\cite{kamioka1,kamioka2,sno3,sno4}, and liquid scintillator experiments~\cite{bore1,bore2,bore3,bore4,bore5,kamland2,kamland3} have not yet demonstrated  the upturn of the recoil energy spectrum
 expected from the MSW effect. 
 This null observation has motivated several theoretical explanations including models with sterile neutrinos~\cite{sterile1,sterile2,sterile3}, mass-varying neutrinos~\cite{massvary}, and non-standard interactions~\cite{non1,non2}.

At energies below $5$~MeV Super-K has observed an excess of events close to the detector structure near the bottom and barrel regions of the detector~\cite{sk4_solar}. However, due to the limited energy resolution of the detector~\cite{Rn_sk}, this energy region overlaps with that of the primary background due to electrons from the $\beta$ decay of $\mathrm{^{214}Bi}$~(Rn daughter). The $Q$-value of this decay is $E_{\beta}^{\mathrm{max}}=3.27$~MeV. 
 In order to understand this background and its contribution to the analysis of solar neutrinos, 
precise measurements of the Rn concentrations in the water of SK-IV detector and in the outputs of the system are necessary.

\subsection{Water purification system}
\label{sec_1_2}

\begin{figure}[]
\centering
\includegraphics[width=100mm]{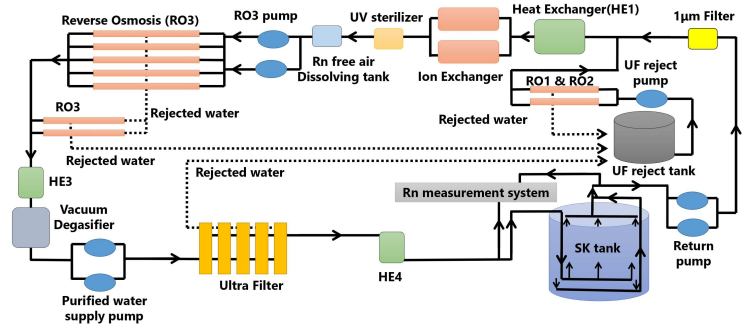}
\caption{Schematic diagram of the SK-IV water system. Purified water is supplied through inlets at the bottom and drained from outlets in both the bottom and top regions as shown in details in Fig.~\ref{flow_bot}.}
\label{water_system}
\end{figure}

The original water purification system for Super-K has been described in Ref.~\cite{Radon_takeuchi_1999}.
During the first phase of Super-K~(SK-I), it consisted of mechanical $1~\mu\mathrm{m}$ filters, a heat exchanger~(HE1), mixed-bed de-ionization resins~(Ion Exchanger,~IE), ultraviolet~(UV) sterilizers, a vacuum degasifier~(VD), a high-quality IE~(Cartridge Polisher,~CP), and ultrafilters~(UF). Their ordering here is the same as that of their appearance in the recirculation process.
Over the intervening $20$~years, the system has been continuously upgraded to reduce impurities and improve control over 
the detector environment. 
 For example, a reverse-osmosis membrane~(RO) was added to the recirculation line during the second phase of operations (SK-II), 
and an additional heat exchanger~(HE3) was installed during the third phase~(SK-III).
 Furthermore, at the beginning of SK-IV an upgraded heat exchanger (HE4) was added to system in order to further control the input water temperature and thereby 
suppress convection in the tank water, which causes Rn and other impurities to mix into the fiducial region of the detector (details below).
Afterwards the water temperature is controlled at $13.06\mathrm{^{\circ}C}$ with an accuracy of $0.01\mathrm{^{\circ}C}$~\cite{Calib_p,sekiya}.
The total circulation rate during SK-IV was $60$~ton/h, which doubles that of SK-III.
Fig.~\ref{water_system} shows the configuration of the system at the end of SK-IV.

\begin{figure}[]
\centering
\includegraphics[width=130mm]{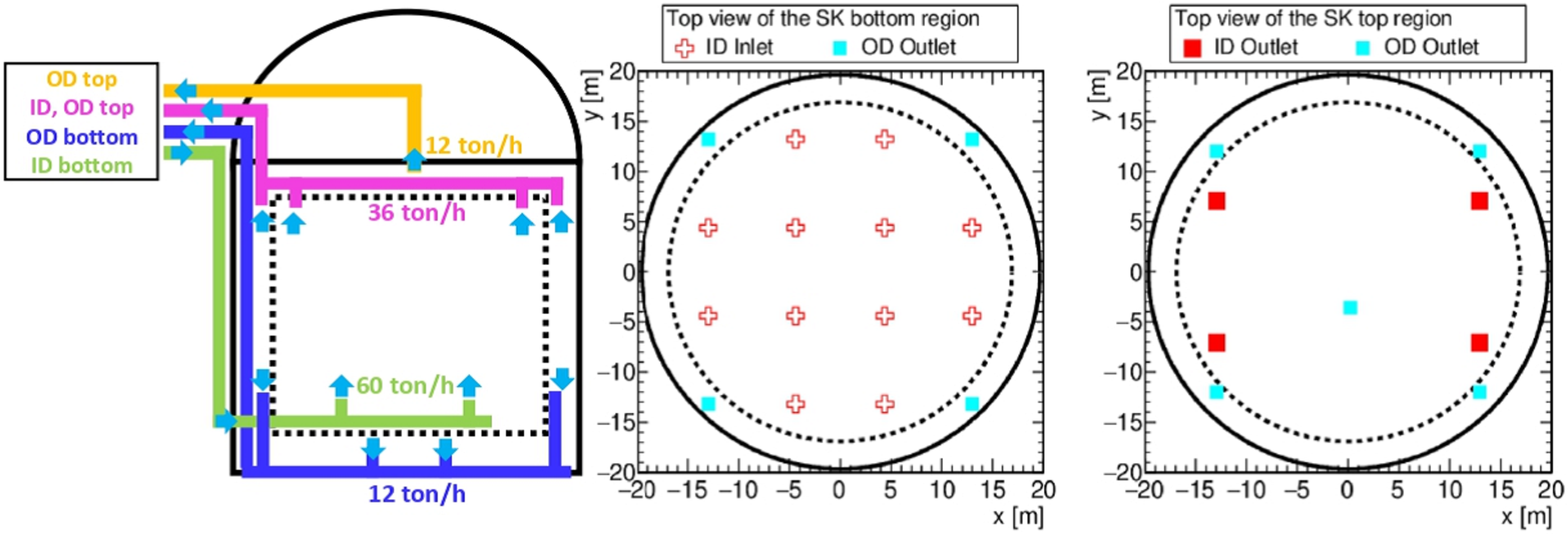}
\caption{Left: Schematic view of the water circulation paths in the Super-K tank. The solid line shows the Super-K water tank, and the dashed line shows the inner detector. The arrows show the direction of water flows. Middle and Right: The locations of the water inlets and outlets at the bottom~(top) of the Super-K tank. The $x$ and $y$ coordinate axes are drawn as defined in Ref.~\cite{paper_SK}. The middle figure illustrates their locations in the bottom region, and the right figure illustrates their locations in the top region. The solid circles describe the Super-K water tank, while the dashed circles outline the inner detector in the tank. The outlets placed in the OD region drain the water from the ``OD barrel'' region.}
\label{flow_bot}
\end{figure}

Water is supplied to Super-K through inlets at the bottom region of the inner detector~(ID), which is shown in Fig.~\ref{flow_bot}.
The inlets extend up to $z=-16.5$~m in the tank, which is $40$~cm below the bottom edge of the fiducial volume used for analysis~\cite{sk4_solar}.
The ID is separated optically from the outer detector~(OD) by a Tyvek sheet~\cite{Calib_p}, but the water in the tank still flows from ID to OD.
As shown in Fig.~\ref{flow_bot}, water is drained through outlets 
placed at the top region of the tank and at the bottom region of the OD for recirculation and purification. The outlets placed in the OD drain water from its barrel region (``OD barrel'').

There are twelve inlets placed at the bottom of ID and four outlets placed in OD.
 As a result of this configuration and the precise temperature control enabled by  HE4, 
the  water tank above $z=-11$~m experiences laminar flow, and therefore lower Rn backgrounds, while that below this level undergoes convection~\cite{Calib_p}.

\subsection{History of Rn studies at Super-K}
\label{sec_1_3}

Several techniques for evaluating the Rn and radium~($\mathrm{^{226}Ra}$) concentrations in water have been developed for underground experiments~\cite{sno_rn, sno_ra1, sno_ra2, bore_rn1, bore_rn2, dayabay, juno_rn, juno_rn2}.
In particular, a Rn assay system for the Super-K water was developed during SK-I~\cite{Radon_takeuchi_2003}. The sensitivity of that detector is 
$0.7~\mathrm{mBq/m^{3}}$ for a single-day measurement and is limited by statistical fluctuations in the background count rate. 
Using that system, the Rn concentration in the supply water was measured to be $0.4\pm0.2~\mathrm{mBq/m^{3}}$ in 2001, while that in the tank water itself was $<2.0~\mathrm{mBq/m^{3}}$~\cite{paper_SK}. 
This was the last of such measurements prior to the results discussed below.  
In order to understand the Rn concentration sufficiently, we require sensitivity to the Rn concentration in purified water at the  $0.1~\mathrm{mBq/m^{3}}$ level.

\section{Experimental setup}
\label{sec_2}

In order to measure ultra-low levels of Rn in water, Rn must first be extracted into air so that daughters of Rn~(especially, $\mathrm{^{218}Po}$ and $\mathrm{^{214}Po}$) can be electrostatically collected in air because they are tend to have a positive charge~\cite{elect_method}. 
This technique requires efficient Rn extraction as well as trapping to allow enough atoms to be collected during a measurement.
To accomplish this we have developed a new water-air extraction column
and have expanded on the $80$~L electrostatic detection system detailed in~\cite{ynakano,PTEP} by introducing a chilled charcoal trap to enhance the Rn collection. 
The accumulated Rn is released and transferred to a detector, whose sensitivity is typically $0.5~\mathrm{mBq/m^{3}}$ in air for a single-day measurement~\cite{ynakano}.
 Finally, an overall sensitivity of ${\sim}0.1~\mathrm{mBq/m^{3}}$ in water has been achieved as determined by fits to the shape of the Rn decay curve.
This represents an order of magnitude improvement over the previous systems~\cite{Radon_takeuchi_1999, Radon_takeuchi_2003}.

\subsection{Column for radon extraction}

\begin{figure}[t]
\centering
\includegraphics[width=110mm]{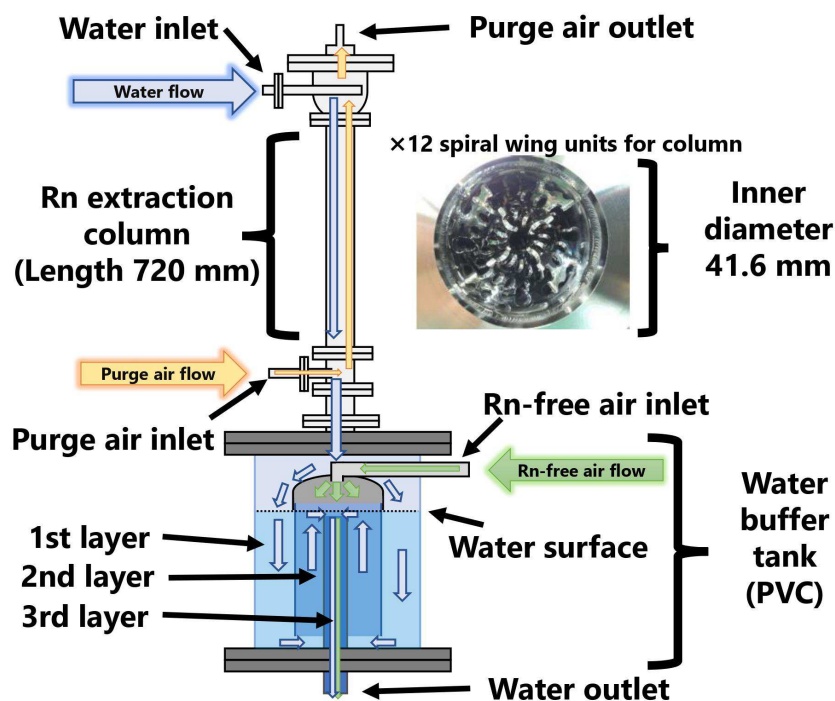}
\caption{Schematic diagram of the Rn extraction system.
 The top~(bottom) part of the system is the extraction column~(water buffer tank).
 Numerical values give lengths in millimeters~(mm), while solid (open) arrows 
 show the flow of sampled water (purge air). 
}
\label{mixer}
\end{figure}

Fig.~\ref{mixer} shows a schematic diagram of the extraction column, which mixes flowing water with purge gas (air) to extract Rn.
 The extraction column consists of $12$~spiral wing units and is equipped with inlets for both the purge air and the sampled water as well as outlets for Rn-degassed water and purge air.
 A buffer tank is located below the column to store Rn-degassed water.

The spiral wing unit\footnote{The unit itself is constructed with a MU-reactor$\mathrm{^{TH}}$, a product of the MU Co., Ltd. http://www.mu-company.com/en\_index.html} 
is a combination of six right-turn units and six left-turn units, which are welded to each other alternately.
 The inside of a wing unit has four welded wings that are directed downward with a height, inner diameter, and outer diameter of 
$60.0$~mm, $41.6$~mm and $48.6$~mm, respectively.
 Each wing contains several holes to improve its Rn extraction efficiency.
 The surfaces of the wings and holes have been electro-polished in order to reduce emanation of Rn from their surfaces. 
 
Degassed water passing through the extraction column is collected in the buffer tank.
 The tank is made of transparent polyvinyl chloride~(PVC), so the water level can be monitored.
 In order to prevent gasses from the external environment from entering into the extraction column, 
the tank is divided into three layers~($1$st, $2$nd and $3$rd layers shown in Fig.~\ref{mixer}) and Rn-free air is supplied to the innermost 3rd layer during operations. The degassed water is finally drained from the pipe of water outlet~(the 3rd layer in Fig.~\ref{mixer}). This design allows the extraction column to maintain both the water level and the inner pressure at the 1st layer. 

Urethane gaskets have been used between the extraction column and water tank here since this material was found to emanate less Rn~\cite{sekiya,bore_rn2} though EPDM~(ethylene propylene diene monomer) or butyl gaskets are commonly used to connect pipes in such systems.

Sampled water enters through the water inlet at the top and at the same time purge air enters the extraction column through the purge air inlet.
When the sampled water falls down through the extraction column and strikes its wings, the water is turned into 
mist allowing the dissolved Rn to escape into the air and be transported out of the system via the upper air outlet.

The total amount of Rn in the water and in the air during this process is conserved before and after the extraction, as expressed by the following equation: 

\begin{equation}
C_{w,0}F_{w} + C_{a,0}F_{a} = C_{w}F_{w} + C_{a}F_{a}. \label{for_1}
\end{equation}

{\noindent}Here $C_{w,0}$~$(C_{a,0})$ is the Rn concentration of the sampled water~(purge air) before extracting in units of $\mathrm{Bq/L}$, $C_{w}$~$(C_{a})$ is that of degassed water~(purge air) after extracting, and  $F_{w}$~$(F_{a})$ is the flow rate of the water~(air) through the system in units of $\mathrm{L/min}$.

In order to evaluate the Rn extraction efficiency, we rewrite Eq.~(\ref{for_1}) as follows:
\begin{align}
C_{w,0}F_{w}  & = C_{w}F_{w}+(C_{a}-C_{a,0})F_{a}, \notag \\
1 & = \frac{C_{w}}{C_{w,0}}+\frac{C_{a}-C_{a,0}}{C_{w,0}}\times\frac{F_{a}}{F_{w}}. \label{eff_form}
\end{align}
\noindent

The second term on the right side in Eq.~(\ref{eff_form}) can be regarded as the Rn extraction efficiency of the system. Note that the total amount of extracted Rn depends on the flow rates of the sampled water and the purge air.  Accordingly, the factor~($F_{a}/F_{w}$) in the second term is required to normalize  the extraction efficiency by taking into account the total volume of water and air used in the Rn extraction (mixing) process.
It can be determined by measuring the Rn concentrations in both the sampled water and the purge air before and after the extraction process.

\begin{figure}[]
\centering
\includegraphics[width=100mm]{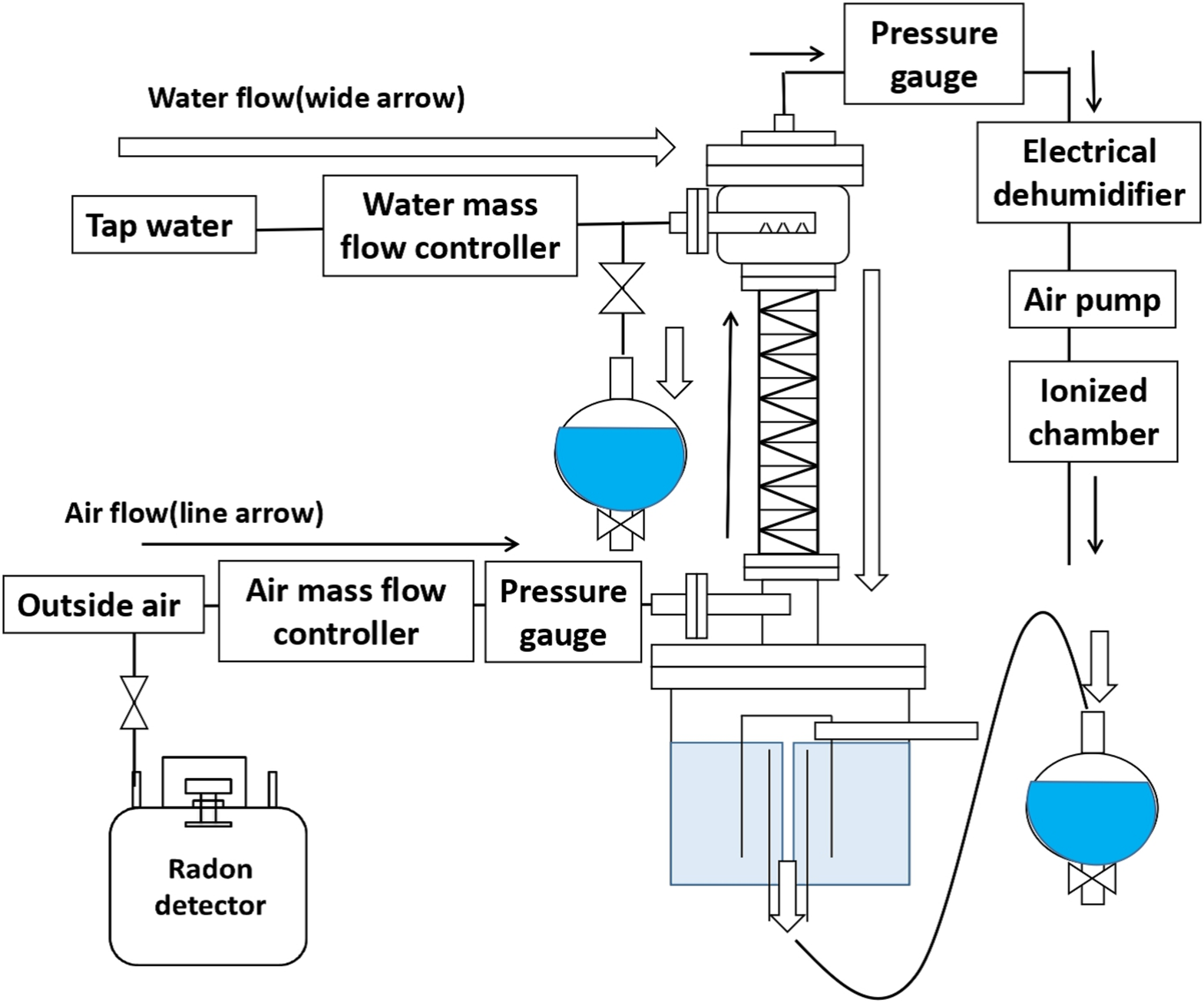}
\caption{Schematic diagram of the calibration setup. Thin arrows show the direction of the air flow, 
while wide arrows show the direction of the water flow.}
\label{calib_mix}
\end{figure}

To determine the Rn extraction efficiency, we built a calibration system at Gifu University as shown in Fig.~\ref{calib_mix}.
The system consists of a $70$~L Rn detector~\cite{Radon_takeuchi_1999}, an air mass-flow controller~(HORIBA STEC Z512), two pressure gauges~(Naganokeiki Co. Ltd. ZT67), a water mass-flow controller~(TOFCO Corp. FLC620), an electrical dehumidifier~(KELK DH-209C), an air pump, and an ionization chamber~(OHKURA ELECTRIC Co. Ltd. RD1210B).

Purge air for the calibration was taken from environment air at Gifu University 
and had a typical Rn concentration of ${\sim}0.01~\mathrm{Bq/L}$ as measured by 
the $70$~L Rn detector.
After the extraction, the concentration in the outflow air was measured with the ionization chamber.
Tap water from the university was used as the Rn source. 
Its Rn content was measured with a liquid scintillator counter~(LSC) system~(Tri-Crab 2900TR produced by PerkinElmer Inc.),
as is standard for evaluation of hot spring water~\cite{lsc3, lsc2, lsc}, and was found to be $5$--$7~\mathrm{Bq/L}$.
The pressure inside the extraction column was monitored by two pressure gauges, located at the inlet and the outlet of the extraction column, 
because the Rn extraction efficiency may depend on this quantity.
The electric dehumidifier was installed just after the extraction column to remove water vapor from the output air 
before sending it into the ionization chamber.
Using the mass-flow controller, the water flow rate was maintained at either $F_{w}=4.0~\mathrm{L/min}$ or $F_{w}=3.58~\mathrm{L/min}$;
which  are flow rates used in  the Super-K measurements described in Section~\ref{sec_measure}.

Calibrations were performed as follows. 
First, tap water and purge air are supplied to the extraction column. After filling the buffer tank with the degassed water, we controlled their flow rates to maintain a constant water level in the PVC vessel~($1$st layer in Fig.~\ref{mixer}). We then sampled the tap water and the Rn-degassed water at the same time using shake flasks. 
Note that when we found air bubbles on the inner surface of the shake flasks they were carefully removed 
in order to avoid Rn exchange between the sampling water and the air-bubbles.
During the extraction process, the Rn concentrations in the purge air and output air were monitored 
using the $70$~L Rn detector and the ionization chamber.
Finally, the Rn extraction efficiency was determined according to Eq.~(\ref{for_1}).

The sources of systematic uncertainties in the calibration measurements are summarized in Table~\ref{sys_calib} and 
are primarily based on estimations from the technical specifications of the measurement devices. 
Measurements of the same sampling vials used in calibration found fluctuations in the Rn concentration of $\pm7.5\%$~\cite{lsc}, which have been attributed to potential air leaks.
If we assume that air leaks occurred during our measurements, the Rn concentration in the water may decrease, 
resulting in an erroneously low measured value.
This would lead to an erroneously high extraction efficiency.
In order to compensate for such problems, we take the value of these fluctuations as a systematic uncertainty. 
Additional systematic uncertainties are taken on the water flow rate stability, $\pm2.0\%$, and the stability of the air flow rate, which is also $\pm2.0\%$.

\begin{table}[]
\caption{Systematic uncertainties in the calibration of the extraction efficiency. The second column shows typical values 
measured during the calibration procedure described in the text.}
\begin{center}
\begin{tabular}{ccc} 
\hline
 Systematic uncertainty & Typical value & Estimated uncertainty \\ \hline 
Ionization chamber & $8$--$12$~Bq/L & $\pm5.0\%$ \\ 
70~L Rn detector~\cite{Radon_takeuchi_1999}& $0.01$~Bq/L & $\pm6.8\%$ \\ 
Liquid scintillation counter & $7$--$9$~Bq/L or $2$--$3$~Bq/L & $\pm10.0\%$ \\
Air leaks in sampling vials & -- & $\pm7.5\%$ \\
Water flow rate & $3.58$~L/min or $4.0$~L/min & $\pm2.0\%$ \\ 
Air flow rate & $2.0$~L/min & $\pm2.0\%$ \\ \hline 
\end{tabular}
\end{center}
\label{sys_calib}
\end{table}

\begin{figure}[]
 \begin{center}
 \includegraphics[width=100mm]{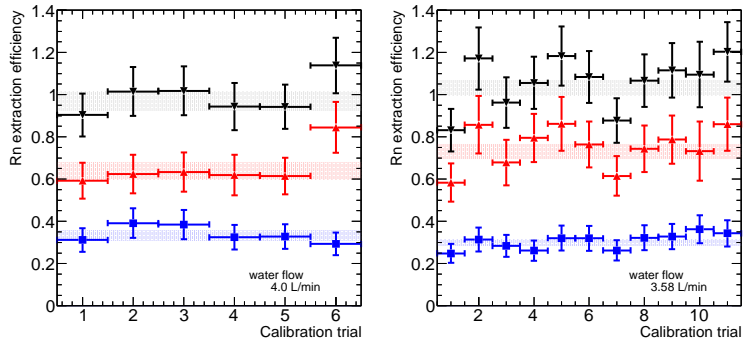}
 \end{center}
\caption{Efficiencies of the Rn extraction from the calibration measurements. The left figure shows the results for the water flow rate of $4$~L/min and the right figure shows results for $3.58$~L/min. Blue squares~(red upward-pointing triangles) show the first~(second) term defined in Eq.~(\ref{eff_form}), and black downward-pointing triangles show their sum. 
Shaded bands show the total uncertainties for each term.
}
\label{mixer_result}
\end{figure}

In total, we performed $6$~($11$) calibrations with the water flow rate set to $F_{w}=4.0$~L/min ($F_{w}=3.58$~L/min), as shown in Fig.~\ref{mixer_result}.
The measured extraction efficiencies are summarized in Table~\ref{table_ext}.
Combining the first term with the second term in Eq.~(\ref{eff_form}), 
we obtain the value of the sum to be $0.97\pm0.05$~($1.03\pm0.04$) for $F_{w}=4.0$~L/min ($F_{w}=3.58$~L/min).
Both values are consistent with $1.0$.
This result demonstrates that the total radioactivity before and after extraction is conserved to within the measurement uncertainty.

\begin{table}[]
\caption{Summary of the measured extraction efficiencies with a constant air flow rate of $F_{a}=2.0$~[L/min]. The first and the second terms are defined in Eq.~(\ref{eff_form}).}
\begin{center}
\begin{tabular}{cc|ccc} 
\hline
$F_{w}$ [L/min] & $F_{a}$ [L/min] &First term & Second term & Total \\ \hline 
$4.00$ & $2.0$ & $0.33\pm0.02$ & $0.64\pm0.04$ & $0.97\pm0.05$ \\ 
$3.58$ & $2.0$ & $0.30\pm0.02$ & $0.73\pm0.03$ & $1.03\pm0.04$ \\ \hline 
\end{tabular}
\end{center}
\label{table_ext}
\end{table}

\begin{figure}[]
 \begin{center}
 \includegraphics[width=100mm]{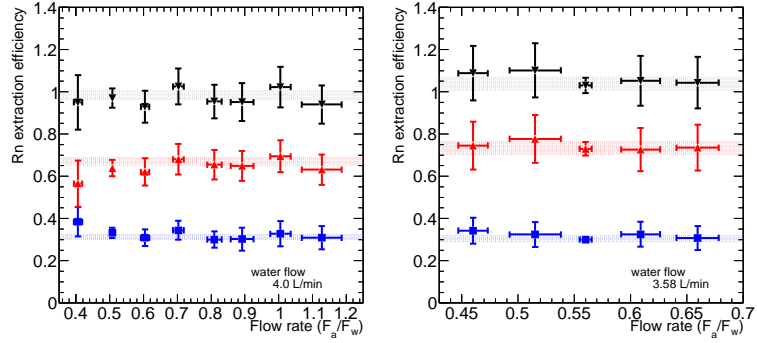}
 \end{center}
 \caption{Dependence of the Rn extraction efficiency on the air flow rate.  
The water flow was fixed at $4.0$~L/min for the left figure and $3.58$~L/min for the right figure, while the air-flow rate was changed from $1.6$~L/min to $4.4$~L/min~(from $1.65$~L/min to $2.35$~L/min).  In the right panel, the third value is calculated by taking average of $11$ measurements shown in Fig.~\ref{mixer_result}. Other values are calculated by using the result from one measurement. Colored markers have the same definitions as in Fig.~\ref{mixer_result}.}
 \label{extraction_of_mixer}
\end{figure}

To understand the stability of the extraction efficiency, we performed additional calibrations 
at a constant water flow of $F_{w}=4.0$~$\mathrm{L/min}$ ($F_{w}=3.58$~L/min) while changing the 
air flow from $1.6$~$\mathrm{L/min}$ to $4.4$~$\mathrm{L/min}$ (from $1.65$ to $2.35$~$\mathrm{L/min}$).
The ratio of the air and water flow rates varies between $0.4$ and $1.1$ (between $0.45$ and $0.65$).
Fig.~\ref{extraction_of_mixer} shows the dependence of the extraction efficiency on this ratio.
There is no correlation as summarized in Table~\ref{table_flow_dep}.

\begin{table}[]
\caption{Air flow rate dependence of the extraction efficiency. The first and the second terms are defined in Eq.~(\ref{eff_form}).}
\begin{center}
\begin{tabular}{cc|ccc} 
\hline
$F_{w}$ [L/min] & $F_{a}$ [L/min] &First term & Second term & Total \\ \hline 
$4.00$ & $1.6$--$4.4$ & $0.31\pm0.01$ & $0.67\pm0.02$ & $0.98\pm0.02$ \\ 
$3.58$ & $1.65$--$2.35$ & $0.31\pm0.01$ & $0.73\pm0.03$ & $1.04\pm0.03$ \\ \hline 
\end{tabular}
\end{center}
\label{table_flow_dep}
\end{table}

%% rvw 
%%
\subsection{Experimental setup and a method for radon concentration measurements}

Charcoal efficiently absorbs various impurities~\cite{maurer} 
and is widely used to trap Rn from several gases~\cite{paper_SK, ynakano, sno_rn, hardy, xmass, lux, motoyasu}.
It has almost $100\%$ trapping efficiency below $-60\mathrm{^{\circ}C}$ and the trapped Rn can be removed with ${\sim}100\%$ efficiency 
by heating the charcoal upto ${+}120\mathrm{^{\circ}C}$ \cite{charcoal}. 
In order to take an advantage these properties, we designed a simple trap using a charcoal-filled $1/2$~inch 
U-shaped electro-polished stainless steel pipe.
The trap was filled with $12.5$~g of charcoal (DIASORB G4-8, produced by Calgon Carbon Japan KK).
This charcoal has also been used in previous studies~\cite{motoyasu}.
When trapping Rn, the U-shaped pipe is placed in a refrigerated ethanol bath.
To release Rn from the trap, the U-shaped pipe is removed from the bath and heated with a band heater. 

\begin{figure}[t]
\centering
\includegraphics[width=100mm]{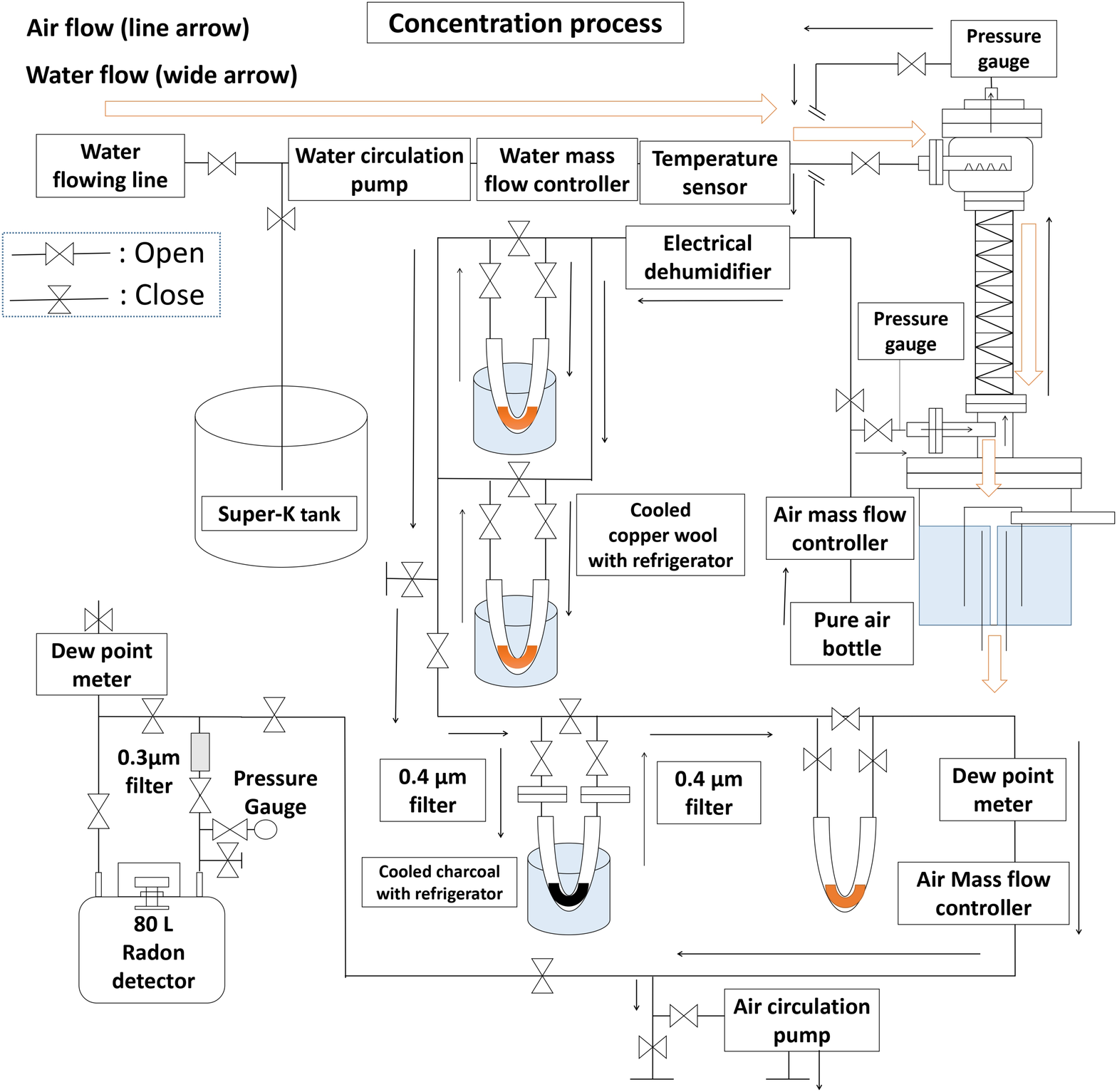}
\caption{Schematic diagram of the setup for the Rn concentration process. 
The thin arrows show the direction of the purge air flow, while thick arrows show the direction of the sampled~(and degassed) water.}
\label{sys_absorption}
\end{figure}

Fig.~\ref{sys_absorption} shows a schematic diagram of the entire Rn measurement system.
It consists of a water pump~(Iwaki Co.  Ltd.,  MDG-R15T100), 
a water mass-flow controller~(TOFLO Corp., FLC620), a temperature sensor~(TOFLO Corp., CF-SCMT, PTM-23), 
the Rn extraction system described in Section~\ref{sec_2}, two pressure gauges~(Naganokeiki Co. Ltd., ZT67), 
an electrical dehumidifier~(KELK, DH-209C), 
three copper wool traps for further water removal,
%%with a refrigerator, 
the charcoal trap, a dew-point meter~(VAISALA, DMT340), 
an air mass-flow controller~(HORIBA STEC, Z512), an air-circulation pump, and the $80$~L Rn detectors~\cite{ynakano, PTEP}.
We used three Rn detectors to conduct three measurements in parallel and they are identical.

%% rvw
A water sample can be supplied from the Super-K tank, from its pure water supply line, or from the return line to the water purification system. 
We used commercially-available G1-grade high-purity air (impurity $<0.1$~ppm) as the purge gas to minimize intrinsic Rn backgrounds.
It is important to remove water from the air before both the charcoal trap and the Rn detector, as their efficiencies depend on the humidity~\cite{humid}.
The electrical dehumidifier and three copper wool traps are used for this purpose. 
Each copper wool trap is a $3/4$~inch U-shaped pipe filled with $12.5$~g of $\varphi~80~\mu\mathrm{m}$ copper wool~(Nippon Steel Wool Cp., Ltd.).
We placed these in an ethanol bath kept below $-80\mathrm{^{\circ}C}$.
Note that Rn is not captured by the cooper wool\footnote{We evaluated the capture rate of Rn on the copper wool with another calibration set up. We confirmed that the Rn concentration did not decrease when the pure air was circulated through the cooled copper trap in the closed set up. Thereby, the capture of Rn on the copper wool is negligible in the analysis described in Section~\ref{sec_23}.}.
Further, we installed $0.4$~$\mu\mathrm{m}$ mesh filters~(Pall Corp., CNF1004USG6) 
before and after the Rn trap to prevent pieces of charcoal from escaping into the measurement system. 
We used $1/2$ inch nylon tubes~(NITAA, MOORE N2) to sample the water.
Other system components are made of electro-polished stainless steel (NISSHO Astec Co., Ltd., MGS-EP SUS316L)
and all joints are connected by VCR$^{\textcircled{\footnotesize{R}}}$ gaskets to minimize Rn emanation and possible air leaks, both of which can affect backgrounds in the measurement.

There are three steps to measure the Rn concentration in water. 
First, Rn is extracted from the water using the extraction column described before and is concentrated in the chilled charcoal trap (concentration process).  
In the next step, the Rn gas is released from the trap and transferred to the Rn detector (transfer process).
Finally, the Rn concentration is measured with the $80$~L detector (measurement process).

Before any measurement, the entire system except for the water lines, the extraction column, and the charcoal trap is first evacuated down to $<1.0\times10^{-4}$~Pa.
The air leak rate of the system was measured to be less than $10^{-10}~\mathrm{Pa\cdot m^{3}/sec}$ 
using a helium leak detector (HELIOT 712D2, produced by ULVAC Equipment Sales Inc).
 The trap is then heated to $+200\mathrm{^{\circ}C}$ for about one hour to completely remove any residual Rn.
Afterwards the system is filled to atmospheric pressure with commercially-available G1-grade high purity air~(impurity concentration $<0.1$~ppm) and the trap is cooled with a refrigerator. 
During the concentration process, we set the water sampling rate to $3.58~\mathrm{L/min}$ or $4.0~\mathrm{L/min}$ and set the flow rate of the purge air to $2.0~\mathrm{L/min}$.
After the water level and the air pressure in the extraction column stabilize, valves before and after the charcoal trap were opened, as shown in Fig.~\ref{sys_absorption}. Sampling and concentration periods varied from $0.5$~hours to $18$~hours, depending upon the expected Rn concentration of a given measurement.

After the concentration process, the valves before and after the charcoal trap were closed and the water sampling was stopped. During the transfer process, we heated the trap to $+200\mathrm{^{\circ}C}$ and 
then opened the valves again to supply pure air at $1.0~\mathrm{L/min}$ in order to fill the Rn detector with the accumulated Rn.
The entrance valve was closed after the Rn detector reached atmospheric pressure.
After these procedures measurements were started and performed typically for $20$~days  in order to determine the shape of the Rn decay curve.

%% rvw 
\subsection{Analysis method}
\label{sec_23}

\begin{figure}[]
 \begin{center}
 \includegraphics[width=100mm]{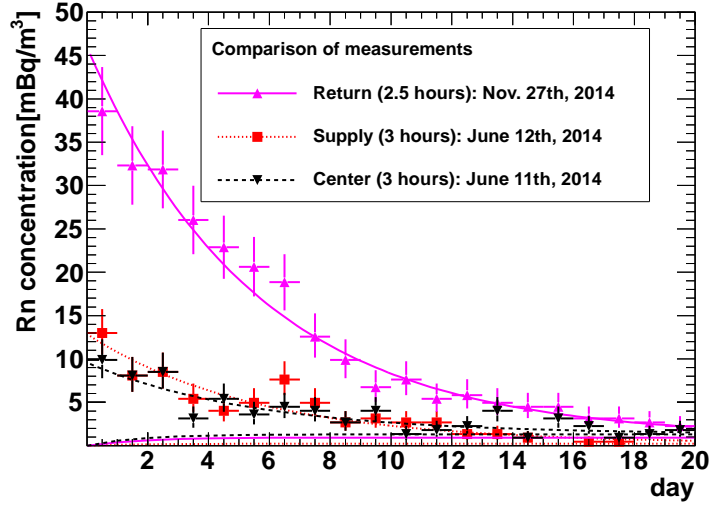}
 \end{center}
 \caption{Measurement results. Super-K detector supply water Rn concentration~(red), that of its return water~(pink), 
and water from its center~(black). The horizontal axis shows the elapsed time and the 
vertical axis shows the Rn concentration as measured with the $80$~L Rn detector. 
The fitted curves are defined by Eq.~(\ref{fit}).}
 \label{results_input}
\end{figure}

Since Rn decays with a constant half-life its initial concentration at the start of the measurement can be derived from the measured decay curve. 
Fig.~\ref{results_input} shows an example of the measured concentration as a function of time.
The data is fitted with the following function:
\begin{equation}
C(t) = Ae^{-\lambda t}+B_{\mathrm{detector}}~(1-e^{-\lambda t}), \label{fit}
\end{equation}
\noindent where $t$~[day], $C(t)$~[$\mathrm{mBq/m^{3}}$], $\lambda$~[$\mathrm{day^{-1}}$], $A$~[$\mathrm{mBq/m^{3}}$], and $B_{\mathrm{detector}}~[\mathrm{mBq/m^{3}}]$ are the elapsed time since the start of the measurement, the Rn decay constant~($\lambda=\ln2/\{3.82~\mathrm{day}\}=0.181~\mathrm{day^{-1}}$), 
the Rn concentration at time $t$, the initial Rn concentration, 
and the background due to the activity of $\mathrm{^{226}Rn}$ in the Rn detector, respectively.
These parameters are listed in Table~\ref{table_parameter}.
Note that in a previous publication~\cite{ynakano} we found $B_{\mathrm{detector}}$ to be $0.33\pm0.07$~$\mathrm{mBq/m^{3}}$ for a particular $80$~L detector. After the actual  measurements described in Section~\ref{sec_measure}, we have evaluated the background of three Rn detectors used in the measurement system. The backgrounds were measured by closing the Rn detector. We used about $89$~days data taken from November 1st 2015 to January 29th 2016.  Each backgrounds are $1.24\pm0.10~\mathrm{mBq/m^{3}}$, $0.27\pm0.04~\mathrm{mBq/m^{3}}$, and $0.23\pm0.04~\mathrm{mBq/m^{3}}$, respectively. The first detector has the larger background than that of others. This is because this detector was reused from the previous Rn detector by replacing its top flange and by performing the electropolish to its inner surface. Due to such additional production processes, it was slightly contaminated.

In the fitting process of the analysis, the parameter of the detector background, $B_{\mathrm{detector}}$, is fitted by limiting its value within $\pm3\sigma$ of the statistical uncertainty of each average background rate listed in Table~\ref{table_parameter}.

\begin{table}[]
\small{
\caption{Summary of the parameters used to calculate the Rn concentration in water.}
\begin{center}
\begin{tabular}{cccc}
\hline
Parameter & Value & Unit & Definition \\ \hline
$t$ & & day & Elapsed time since the start of measurement \\ 
$C(t)$ & & $\mathrm{mBq/m^{3}}$ & Rn concentration measured by the Rn detector at time $t$ \\
$A$ & & $\mathrm{mBq/m^{3}}$ & Rn concentration at the start of measurement \\
$\lambda$ & $\ln2/3.82$ & $\mathrm{day}^{-1}$ & $\mathrm{^{222}Rn}$ decay constant \\ \hline
$B_{\mathrm{detector}}$ & $1.24\pm0.10$ & $\mathrm{mBq/m^{3}}$ & Background of $80$~L Rn detector No.~1 \\ 
 & $0.24\pm0.04$ & & Background of $80$~L Rn detector No.~2 \\
 & $0.27\pm0.04$ & & Background of $80$~L Rn detector No.~3 \\ \hline
$t_{\mathrm{con}}$ & & hour & Duration of the concentration process \\ 
$t_{\mathrm{total}}$ & & hour & Combined duration of the concentration and transfer processes \\ 
$F_{a}~(F_{w})$ & & L/min & Purge air~(sampled water) flow rate \\
$C_{\mathrm{PAD}}$ & & $\mathrm{mBq/m^{3}}$ & Rn concentration in purge air after degassing \\
$V_{\mathrm{det}}$ & $0.080$ & $\mathrm{m^{3}}$ & Volume of the $80$~L Rn detector \\ 
$V_{\mathrm{purge}}$ & $F_{a}\times t_{\mathrm{con}}$ & $\mathrm{m^{3}}$ & Purge air volume \\ 
$\beta_{\mathrm{corr}}$ & $\exp(-\lambda t_{\mathrm{total}})$ & & Correction factor due to Rn decay \\
& & & during concentration and transfer processes \\ \hline
$\varepsilon_{\mathrm{mixing}}$ & See Table~\ref{table_ext} & & Extraction column Rn extraction efficiency \\
$\varepsilon_{\mathrm{trap}}$ & $0.99\pm0.01$ & & Charcoal trap trapping efficiency \\ 
$\varepsilon_{\mathrm{rel}}$ & $0.99\pm0.01$ & & Charcoal trap release efficiency \\ \hline
\end{tabular}
\end{center}
\label{table_parameter}
}
\end{table}

%%%%
%%
%  rvw 
%%
%%%%

After obtaining $A$ from the fit, we derived the Rn concentration in the air.
We obtained the initial total radioactivity in the detector as $A\times V_{\mathrm{det}}$, where $V_{\mathrm{det}}=0.080~\mathrm{m^{3}}$ is the total volume of the Rn detector.
On the other hand, the total radioactivity in the purge air after extracting is given by $C_{\mathrm{PAD}}V_{\mathrm{purging}}$, where $C_{\mathrm{PAD}}$ is the Rn concentration in the purge air after the extraction and $V_{\mathrm{purging}}=F_{a}\times t_{\mathrm{con}}$ is the total volume of the purge air that passes through the  charcoal trap during the concentration process~($F_{a}$ is the flow rate of the purge air).
Since the initial total radioactivity $AV_{\mathrm{det}}$ should be the same as $C_{\mathrm{PAD}}V_{\mathrm{purge}}$, 
we obtain the Rn concentration in the purge air after degassing as $C_{\mathrm{PAD}} = AV_{\mathrm{det}}/V_{\mathrm{purge}}$.
Here, the efficiencies of Rn trapping~($\varepsilon_{\mathrm{trap}}$) and release~($\varepsilon_{\mathrm{rel}}$) are assumed to be $0.99\pm0.01$, since the former (latter) reaches almost $100\%$ when the trap is cooled down to $-60\mathrm{^{\circ}C}$~(baked at more than $+120\mathrm{^{\circ}C}$)~\cite{charcoal}, as mentioned above.
Then, we derived the Rn concentration in the sampled water by dividing $C_{\mathrm{PAD}}$ by the Rn extraction efficiency.
Since the concentration and transfer processes take hours, Rn decay during the entire measurement must also be considered.
We used the correction term $\beta_{\mathrm{corr}}=\exp(-\lambda t_\mathrm{total})$ for this analysis, where $t_{\mathrm{total}}$ is the total time required for both the concentration and transfer processes.
Finally, we obtain the Rn concentration of the sampled water ($C_{\mathrm{sample}}$) from
\begin{equation}
C_{\mathrm{sample}} = \frac{A}{\beta_{\mathrm{corr}}}\times\frac{V_{\mathrm{det}}}{V_{\mathrm{purge}}}\times\frac{F_{a}}{\varepsilon_{\mathrm{mixing}}F_{w}}\times \frac{1}{\varepsilon_{\mathrm{trap}} \varepsilon_{\mathrm{rel}}}, \label{form_c}
\end{equation}
\noindent where the parameters used in this equation are listed in Table~\ref{table_parameter}.

\section{Backgrounds and systematic uncertainties}
\label{sec_bg}

Although all materials and system components were  carefully selected, a Rn background still exists due to air leaks and emanation from contaminants on the inner surfaces of the setup.
To evaluate this background we performed several dedicated measurements.
Note that this background, which we define as $B_{\mathrm{system}}$, is to be 
distinguished from the intrinsic background of the Rn detector itself, defined as $B_{\mathrm{detector}}$ in Eq.~(\ref{fit}).
Possible background sources are (A)~residual Rn in the G1-grade pure air,
(B)~Rn emanation from the inner surface of the air flow line, including the electrical dehumidifier and copper wool traps,
(C)~emanation from the inner surface of the extraction column, including the PVC acrylic vessel,
(D)~emanation from the charcoal,
and (E)~emanation from the components of the water sampling line, such as the nylon tube and the water pump.
The total background in the system is given by $B_{\mathrm{system}} = B_{\mathrm{air-line}}+B_{\mathrm{column}}+B_{\mathrm{water-line}}$, where $B_{\mathrm{air-line}}$ is the backgrounds  (A), (B) and (D), 
$B_{\mathrm{column}}$ is the background from (C), 
and $B_{\mathrm{water-line}}$ is that from (E).

\subsection{Backgrounds from the air lines}

In order to evaluate $B_{\mathrm{air-line}}$, we performed measurements after bypassing the extraction column.
The valves before and after the extraction column were closed and air was supplied directly to the dehumidifier, copper wool traps, and the charcoal trap during the concentration process.
We then performed the transfer and measurement processes and obtained $B_{\mathrm{air-line}}$ from
\begin{equation}
B_{\mathrm{air-line}}= \frac{A}{\beta_{\mathrm{corr}}}\times\frac{V_{\mathrm{det}}}{V_{\mathrm{purge}}}\times \frac{1}{\varepsilon_{\mathrm{trap}} \varepsilon_{\mathrm{rel}}}, \label{form_bg}
\end{equation}
\noindent where $A$, $\beta_{\mathrm{corr}}$, $V_{\mathrm{det}}$, $V_{\mathrm{purge}}$, $\varepsilon_{\mathrm{rel}}$ and $\varepsilon_{\mathrm{trap}}$ are listed in Table~\ref{table_parameter}.

{\noindent} We performed four measurements for different durations of the concentration process~($19$, $21.5$, $22$, and $24$~hours).
During the concentration process of these measurements, we cooled the charcoal trap down to $-70\mathrm{^{\circ}C}$ as usual. After this process, we baked the charcoal trap to release the absorbed Rn and transferred the accumulated Rn to the Rn detector as usual. 
These measurements give the total background caused by intrinsic background from purified air, Rn emanation from the charcoal, and from the system itself. 
Using Eq.~(\ref{form_bg}) we estimated the backgrounds from the air line to be $B_{\mathrm{air-line}}=0.013\pm0.002~\mathrm{mBq/hour}$, which is also listed in Table~\ref{table_slope_bg}. 
Since this background is proportional to some extent to the total duration of the concentration process~($t_{\mathrm{con}}$), this estimated background can be normalized by the total amount of air used 
and can be expressed as $B_{\mathrm{air-line}}=0.11\pm0.02~\mathrm{mBq/m^{3}}$. 
Accordingly, the intrinsic Rn concentration in the G1-grade pure air is below this level.

\begin{table}[]
\caption{Summary of background evaluated in Section~\ref{sec_bg}. As described in the main text, the total background is proportional to the total duration of measurement processes. Therefore, the hourly background is listed in the second column. In addition to this, the Rn concentration of background is also calculated by normalizing  total amount of used air~(sampled water) as listed in the third column.}
\begin{center}
\begin{tabular}{ccc} 
\hline
Background &  $\mathrm{mBq/hour}$ & $\mathrm{mBq/m^{3}}$ \\ \hline 
$B_{\mathrm{air-line}}$  & $0.013\pm0.002$ & $0.11\pm0.01$ \\
$B_{\mathrm{column}}+B_{\mathrm{water-line}}$  & $0.143\pm0.027$ & $0.67\pm0.12$ \\ \hline
\end{tabular}
\end{center}
\label{table_slope_bg}
\end{table}

%%
%%  RVW
%%  RVW
%%  RVW
%%

\subsection{Backgrounds from the extraction column and water lines}
 
In order to determine $B_{\mathrm{mixer}}+B_{\mathrm{water-line}}$, a closed-loop running~\cite{sno_rn, bore_rn2}, where Rn-degassed water is sent to the extraction column repeatedly, is used.
From Eq.~(\ref{eff_form}), the Rn concentration in the degassed water after one extraction cycle 
is given by $C_{w}=C_{w,0}(1-\varepsilon_{\mathrm{mixing}})=C_{w,0}p$, where $C_{w}$, $C_{w,0}$, and $\varepsilon_{\mathrm{mixing}}$ 
are as defined in Section~\ref{sec_23}.  Here $p$ is $(1-\varepsilon_{\mathrm{mixing}})$.
The Rn concentration in the sampling water just before the extraction column is 
%%%
\begin{equation}
C_{w,0}=C_{\mathrm{sample}}+B_{\mathrm{water-line}}, 
\end{equation}
\noindent where $C_{\mathrm{sample}}$ is the concentration in the sampled water. 
The concentration in the Rn-degassed water is then
\begin{equation}
C_{\mathrm{degassed}} = p(C_{w,0}+B_{\mathrm{column}})=p(C_{\mathrm{sample}}+B_{\mathrm{water-line}}+B_{\mathrm{column}}).
\end{equation}
\noindent After performing the extracting processes $n$ times ~(i.e., after looping through the system $n$ times), 
the Rn concentration in the degassed water~($C_{n\mathrm{-looped}}$) can be written as
\begin{equation}
C_{n\mathrm{-looped}}=p^{n}C_{\mathrm{sample}}+\frac{1-p^{n}}{1-p}(B_{\mathrm{water-line}}+B_{\mathrm{column}}).
\end{equation}
Therefore, after a large number of mixings,
the concentration, $C_{\mathrm{closed-loop}}$, reaches an equilibrium that is determined by the background and the extraction efficiency:
\begin{equation}
C_{\mathrm{closed-loop}}=\frac{1}{1-p}(B_{\mathrm{water-line}}+B_{\mathrm{column}}).
\end{equation}

In order to prepare water for the closed-loop running, we conducted this operation for more than 6 hours~($n>50$) and 
measured the Rn concentration following the method described in the previous section.
The result is $C_{\mathrm{closed-loop}}=0.196\pm0.028~\mathrm{mBq/hour}$ for a water flow rate $F_{w}=3.58$~L/min.
Thus, we obtain the background as $B_{\mathrm{water-line}}+B_{\mathrm{column}} = 0.143\pm0.027~\mathrm{mBq/hour}$ as listed in Table~\ref{table_slope_bg}. 
We use this value in the measurements below.
Moreover, the estimated background is normalized by the total amount of sampled water and 
expressed as $B_{\mathrm{water-line}}+B_{\mathrm{column}} = 0.67\pm0.12~\mathrm{mBq/m^{3}}$, which corresponds to the total background when $1~\mathrm{m^{3}}$ of water is sampled~(about $5$~hours for $3.58$~L/min, bout $4$~hours for $4.0$~L/min.).
Since the inner surface of the extraction column has been electro-polished, the Rn emanation rate from this part of the system is expected to be low.
In contrast, we expect the background from the PVC vessel to be relatively high and it may be the main background source. Systematic uncertainties for measurements with this system are summarized in Table~\ref{sys_meas}.

\begin{table}[]
\caption{Summary of the systematic uncertainties for the measurement system.}
\begin{center}
\begin{tabular}{cc} 
\hline
Source & Systematic uncertainty \\ \hline
Rn extraction efficiency~($\varepsilon_{\mathrm{mixing}}$) & Table~\ref{sys_calib} \\ 
80~L Rn detector calibration& $\pm5.7\%$~\cite{ynakano} \\ 
Difference among three 80~L Rn detectors & $\pm10.0\%$~\cite{PTEP,ynakano} \\ 
Water flow rate~($F_{w}$) & $\pm2.0\%$ \\ 
Air flow rate~($F_{a}$) & $\pm2.0\%$ \\ 
Rn trapping efficiency~($\varepsilon_{\mathrm{trap}}$)& $\pm1.0\%$ \\ 
Rn release efficiency~($\varepsilon_{\mathrm{rel}}$) & $\pm1.0\%$ \\ 
\hline 
\end{tabular}
\end{center}
\label{sys_meas}
\end{table}

\section{Radon concentration measurements}
\label{sec_measure}
\subsection{Supply water}
\label{sec_sup_meas}

\begin{figure}[]
\begin{center}
\includegraphics[width=100mm]{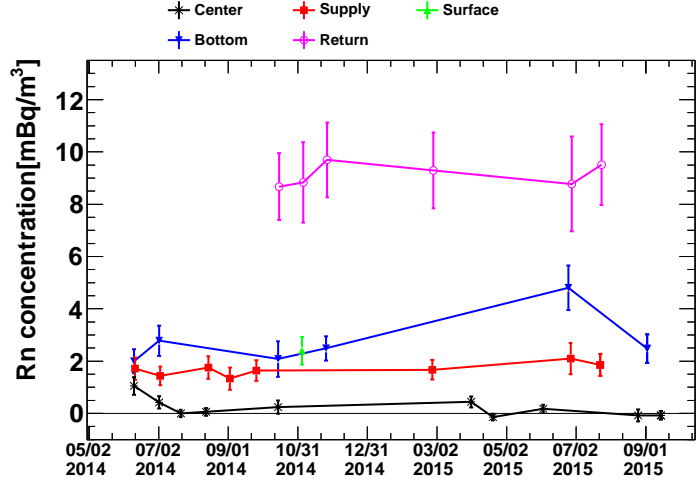}
\end{center}
\caption{Measured Rn concentrations in various Super-K water samples as a function of time.
The black crosses, red squares, green upward-pointing triangles, blue downward-pointing triangles, 
and pink circles show results for the center region, the supply water, water sampled from the surface of the tank water, the bottom region, and the return water, respectively.
}
\label{results_time}
\end{figure}

Since Rn contamination in the Super-K supply water can be a potentially dangerous source of backgrounds, monitoring its concentration is essential to understand those backgrounds and the stability of the detector's response to them.
Measurements of the Rn concentration in the supply water were done using a sampling port located after the last stage of the water 
circulation system~(after HE4 in Fig.~\ref{water_system}).
Using data taken from June 2014 to October 2015, the Rn concentration in the supply water was found to be 
stable at $C_{\mathrm{Supply}}=1.74\pm0.14$~$\mathrm{mBq/m^{3}}$ as shown in Fig.~\ref{results_time}.

\subsection{ID bottom} 

Since water is supplied to the Super-K tank through ID via inlets at its bottom (Fig.~\ref{flow_bot}), 
measurements of the Rn concentration there should track those of the supply water itself.
Water is sampled from this region by inserting a 1/2-inch nylon tube from the top of the detector at 
$(x,y,z)=(+0.353,-0.707)$~m and lowering it to $z=-12.000$~m.
This location is also used to calibrate the detector near its energy threshold~\cite{DT, Calib_p} and  
therefore Rn backgrounds in this area are of particular interest.
We performed several measurements as shown in Fig.~\ref{results_time} 
and found $C_{\mathrm{ID-bottom}}=2.63\pm 0.22$~$\mathrm{mBq/m^{3}}$.
Note that this is a higher concentration than that of the supply water, the implications of which are discussed 
in the next section.

\subsection{Center of the Super-K tank}

Typically the center region of the Super-K detector is the most radio-pure, since it is far from the detector walls, water inputs and returns. 
As a result, it is the best candidate for studying very low energy solar neutrino interactions and consequently
it is essential to understand backgrounds from any residual Rn daughters therein.
Accordingly, water was sampled at the same $(x,y)$ location as discussed above, but at $z=+0.400$~m in order to study the center of the 
detector. 
The results are shown in the thick black line of Fig.~\ref{results_time}.
As expected, the Rn concentration in this region is quite low, often consistent with zero within the measurement errors.
The measurement shows a negative central value, which has been attributed to fluctuations of the background. 
We note that in these cases the upper limit on the concentration is $C_{\mathrm{Center}} <0.23$~$\mathrm{mBq/m^{3}}$~($95\%$~C.L.).

\subsection{Return water}

The SK-IV water circulation system pumps water out of the tank for re-purification via outlets 
placed at the top and bottom of the detector (Fig.~\ref{flow_bot}).
This return water is a mixture of water from the OD surface, the OD barrel~(i.e. the OD outlet shown in the right panel of Fig.~\ref{flow_bot}), 
the ID top, 
and the OD bottom. 
The corresponding flow rates are listed in Table~\ref{table_flow}.
Since a significant fraction of the water is from the OD, which is surrounded by Rn sources, including the tank lining and cavern rock, it is important to measure the return water's Rn concentration to identify potential background sources in the ID.

\begin{table}[]
\caption{Super-K water flow. 
As shown in Fig.~\ref{flow_bot}, the return water is a mixture of the water from the OD surface, ID${+}$OD top, and OD bottom regions.}
\begin{center}
\begin{tabular}{cc} 
\hline
Position & Water flow rate~[$\mathrm{ton/hour}$]\\ \hline
OD surface & 12 \\ 
ID${+}$OD top & 36 \\ 
OD bottom & 12 \\ \hline 
Return & 60 \\ \hline 
\end{tabular}
\end{center}
\label{table_flow}
\end{table}

Return water was sampled from a port located just after the pump to send water from the tank back to the circulation system~(return pump in Fig.~\ref{water_system}).
The measured Rn concentration was $C_{\mathrm{Return}} = 9.06\pm0.58~\mathrm{mBq/m^{3}}$.
It is clear that water that has passed through PMTs and the detector structure has  higher concentration 
than that of the supply water. 
Comparison of the return and supply concentrations indicates that the water circulation system's total Rn removal efficiency 
is $0.81\pm0.08$.

\subsection{Measurement verification} 
\label{sec_prop}
\begin{figure}[]
\begin{center}
\includegraphics[width=120mm]{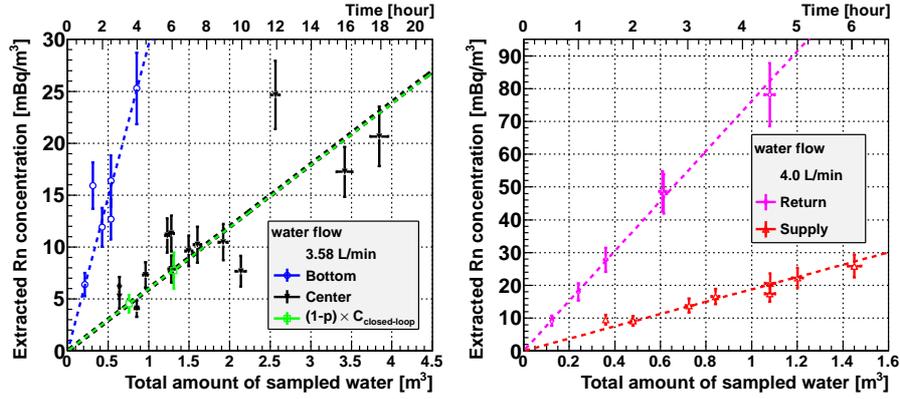}
\end{center}
\caption{Relationship between the extracted Rn concentration and the total amount of sampled water.
The vertical axis shows the extracted Rn concentration after correcting for the time factor $\beta_{\mathrm{corr}}$ 
and the lower~(upper) horizontal axis shows the total amount of sampled water~(total duration of the concentration process, $t_{\mathrm{con}}$ in Table~\ref{table_parameter}).
The left figure shows results for water sampled from the tank, as well as for the closed-loop water, 
at a flow rate $F_{w}=3.58$~L/min.
Similarly, results for the supply and return lines at a flow rate of $F_{w}=4.0$~L/min 
are shown in the right figure.}
\label{volume_vs_con_water}
\end{figure}

As shown in Fig.~\ref{results_time}, the Rn concentrations in various water samples have been stable to within their measurement uncertainties.
Assuming that the concentrations are constant on the time scale of several hours, 
we can test the validity of the measurement as follows.
If the total amount of Rn is proportional to the total amount of sampled water, the accumulated Rn ($A/\beta_{\mathrm{corr}}$) should be a linear function of the total sampled water volume~(integrated flow).
The slope of this function corresponds to the Rn concentration in the sampled water.
Fig.~\ref{volume_vs_con_water} shows the accumulated Rn concentration as a function of the total amount of sampled water 
and displays the expected linear relationship.

By fitting the slope, the Rn concentrations in the sampled water can be derived from the following equation, independent of Eq.~(\ref{form_c}):
\begin{equation}
C_{\mathrm{sampled}} = \frac{V_{\mathrm{det}}\times slope}{\varepsilon_{\mathrm{trap}}\varepsilon_{\mathrm{rel}}\varepsilon_{\mathrm{mixing}}}. \label{form_slope}
\end{equation}
{\noindent} The Rn concentrations obtained with this method are summarized in Table \ref{table_slope} and compared 
with those from Eq.~(\ref{form_c}) after subtracting the backgrounds described in Section~\ref{sec_bg}.
The table also lists the $\chi^{2}$ values and corresponding $p$-values from the fits.
Regarding the latter, we note that with the exception of the center region, which has suffered 
fluctuations in the background, each fit is compatible with the hypothesized linear relationship.
Further, the consistency of results across measurement methods indicates 
that the observed Rn concentrations in the Super-K water are accurate at the $\mathrm{mBq/m^{3}}$ level.

\begin{table}[]
\caption{Summary of the measurement results obtained from the methods of Eq.~(\ref{form_c}) and Eq.~(\ref{form_slope}), 
as well as the  $\chi^{2}$ per degree of freedom and corresponding $p$-value from the latter.}
\begin{center}
\begin{tabular}{cc|ccc} 
\hline
Sampled water & Eq.~(\ref{form_c})& Eq.~(\ref{form_slope}) & $\chi^{2}/\mathrm{d.o.f}$ & $p$-value\\
 & [$\mathrm{mBq/m^{3}}$] & [$\mathrm{mBq/m^{3}}$] & & \\ \hline 
Supply & $1.74\pm0.14$ & $1.64\pm0.50$ & $5.053/7$ & 0.653\\ 
ID bottom & $2.63\pm0.22$ & $2.63\pm0.30$ & $11.68/5$ & 0.039\\ 
Center & $<0.23$~($95\%$~C.L.) & $<0.26$~($95\%$~C.L.) & $40.76/13$ & 0.010 \\ 
Return & $9.06\pm0.58$ & $8.71\pm1.96$ & $0.37/5$ & 0.996 \\ \hline
$(1-p)C_{\mathrm{closed-loop}}$ & $0.67\pm0.12$ & $0.67\pm0.10$ & -- & -- \\ 
$=B_{\mathrm{air-line}}+B_\mathrm{column}$ & & & & \\ \hline
\end{tabular}
\end{center}
\label{table_slope}
\end{table}

\subsection{Comparison with past measurements}

Table~\ref{comp_past} shows a comparison of the Rn concentrations in the supply and return water measured in this study 
as well as previous measurements~\cite{Rn_sk, Radon_takeuchi_1999, Radon_takeuchi_2003}. 
The Rn concentrations in the supply water and return water during SK-IV are higher than those measured during SK-I. 
However, the Rn reduction efficiency of the former ($0.81\pm0.08$) is comparable to that of the latter (${\sim}0.80$). 
Accordingly, the higher Rn concentration in the SK-IV return water resulted in the higher concentration in the supply water.

\begin{table}[]
\caption{Summary of the measured Rn concentration in supply water and return water.}
\begin{center}
\begin{tabular}{cccc} 
\hline
Sampling water  & Phase & Result~$[\mathrm{mBq/m^{3}}]$ & Method\\ \hline
Supply &  Beginning of SK-I & $<3.2$ & 70~L Rn detector w/ plastic ball~\cite{Radon_takeuchi_1999,Rn_sk} \\
Supply &  End of SK-I & $0.4\pm0.2$ & 700~L Rn detector w/ membrane~\cite{paper_SK, Radon_takeuchi_2003} \\ 
Supply &  SK-IV~(This study) & $1.74\pm0.14$ & 80~L Rn detector w/ extraction column  \\ 
\hline
Return &  Beginning of SK-I & $<5.0$ & 70~L Rn detector w/ plastic ball~\cite{Radon_takeuchi_1999,Rn_sk} \\
Return &  End of SK-I & $<2.0$ & 700~L Rn detector w/ membrane~\cite{paper_SK, Radon_takeuchi_2003}  \\
Return &  SK-IV~(This study) & $9.06\pm0.58$ & 80~L Rn detector w/ extraction column \\ \hline
\end{tabular}
\end{center}
\label{comp_past}
\end{table}

In the previous study, the Rn concentrations in the tank were estimated using a Rn injection calibration method~\cite{Rn_sk}. 
Table~\ref{comp_tank} shows a comparison of Rn concentrations in the tank water. 
For the region between $\pm0$~m and $+10$~m an upper limit of $<1.4~\mathrm{mBq/m^{3}}$ was obtained during SK-I. 
Further, the estimated Rn concentration at $z=-6$~m~($z=-11$~m) was $3.0~\mathrm{mBq/m^{3}}$~($5.0~\mathrm{mBq/m^{3}}$). 
On the other hand, the measured Rn concentrations with system used in this manuscript are $<0.23~\mathrm{mBq/m^{3}}$ in the center region~($z=+0.4$~m) and $2.63\pm0.22~\mathrm{mBq/m^{3}}$ in 
the bottom region~($z=-12$~m). Based on those results, the measurement technique presented in the article has improved the measurement sensitivity to the Rn concentration in purified water comparing with the past techniques. 

The results are summarized in Table~\ref{comp_tank} and suggest  that the Rn concentrations in the center region and the bottom region, which are used for solar neutrino analysis, are clearly lower in SK-IV.
Therefore, we conclude that we have successfully reduced the Rn in the both regions by optimizing both the supply water temperature and the water circulation rate. 
On the other hand, this optimization results in higher Rn concentrations in the supply water and in the return water, indicating that the dominant sources of Rn are in the Super-K tank itself and not in the water system or buffer gas (see below).

\begin{table}[]
\caption{Summary of the measured Rn concentration in the tank water.}
\begin{center}
\begin{tabular}{cccc} 
\hline
Detector position  & Phase & Result~$[\mathrm{mBq/m^{3}}]$ & Method\\ \hline
Tank water &  Beginning of  SK-I & $<5.7$ & 70~L Rn detector w/ plastic ball~\cite{Radon_takeuchi_1999,Rn_sk} \\
$0\le z \le 10$~m &  Beginning of  SK-I & $<1.4$ &Rn injection calibration~\cite{Rn_sk} \\
$z=-6$~m &  Beginning of  SK-I & $3.0$ &Rn injection calibration~\cite{Rn_sk} \\
$z=-11$~m &  Beginning of  SK-I & $5.0$ &Rn injection calibration~\cite{Rn_sk} \\ \hline
$z=+0.4$~m &  SK-IV~(This study) & $<0.23$ & 80~L Rn detector w/ extraction column  \\ 
$z=-12$~m &  SK-IV~(This study) & $2.63\pm0.22$ & 80~L Rn detector w/ extraction column  \\ 
 \hline
\end{tabular}
\end{center}
\label{comp_tank}
\end{table}

Further discussion related to Rn sources in the water tank and the observed background rate caused by Rn daughters in the Super-K solar neutrino analysis will be presented in a subsequent publication.

\section{Other OD Measurements}
\label{sec_5}

The bottom region of OD may also have a large Rn concentration because dust from the detector volume may settle and 
accumulate there.
Although measurements of its water may provide hints at possible Rn sources in the tank,
it is impossible to directly sample water from this region. 

However, the Rn concentration in the OD bottom water can be indirectly estimated since part of the 
return water is taken from this region (see Table \ref{table_flow}).
The estimation proceeds via the following equation:
\begin{equation}
C_{\mathrm{OD-bottom}} = \frac{ C_{\mathrm{return}}F_{\mathrm{return}}-\left(C_{\mathrm{OD-surface}}F_{\mathrm{OD-surface}}+C_{\mathrm{ID-top}}F_{\mathrm{ID-top}}+C_{\mathrm{OD-barrel}}F_{\mathrm{OD-barrel}}\right)}{F_{\mathrm{OD-bottom}}},
\end{equation}
{\noindent}where variables beginning with $C$ represent the Rn concentrations of each water sample, 
and similarly, those with $F$ represent the flow rates. 
In order to obtain $C_{\mathrm{OD-barrel}}$ for this estimate, 
we sampled water from the barrel region of the OD barrel by inserting 
a sampling tube at $(x,y,z)=(+17.321,-3.535,+17.000)$~m on November 6, 2014 
and measured the Rn concentration of $3.48\pm0.60~\mathrm{mBq/m^{3}}$.
Although the water tank in the top region of ID has not been measured directly, 
we estimate its Rn concentration to be equal to that of the ID center, i.e., 
$C_{\mathrm{ID-top}}{\sim}C_{\mathrm{center}}$, based on Super-K low energy background data~\cite{sk4_solar, dron, nue2017}.
Using these measurements and assumptions, 
we estimate the concentration in the OD bottom to be $C_{\mathrm{OD-bottom}}=33.97\pm3.30$~$\mathrm{mBq/m^{3}}$.
Note that this region is expected to be the least radio-pure region in the Super-K tank.

In the Super-K tank there is a buffer gas layer between the surface of the OD water and the top of the tank~\cite{paper_SK}.
In previous publications~\cite{ynakano}, we have concluded that Rn contamination from the Super-K tank itself 
is the dominant source of Rn in this  buffer gas.
This suggests that Rn in the buffer gas, whose concentration was measured to be $28.8\pm1.7$~$\mathrm{mBq/m^{3}}$, 
dissolves into the OD top water.
In order to confirm this Rn contamination directly, we measured the Rn concentration at depth of 20~cm below
the surface of the tank water by inserting a nylon tube into the calibration hole at $(x,y)=(-0.950,-1.064)$~m.
The resulting Rn concentration was $2.51\pm0.47$~$\mathrm{mBq/m^{3}}$, indicating that 
the surface water is not the main source of Rn in the buffer gas and suggesting further  the detector structure is the primary Rn source. 

\section{Conclusion and future prospects}
\label{sec_6}

We have developed a new technique for measuring ultra-low Rn concentrations in purified water. 
For this purpose, we developed and calibrated an extraction column to extract Rn from water with 
extraction efficiencies of $0.64\pm0.03$ for a water flow rate of $F_{w}=4.0$~$\mathrm{L/min}$ and 
$0.73\pm0.04$ for $F_{w}=3.58$~$\mathrm{L/min}$ when $F_{a}=2.0$~$\mathrm{L/min}$.
For fixed values of the water flow, we additionally found the efficiency has no dependence on the air flow rate.

Combining the extraction column with an existing 80~L Rn detector 
we have constructed a new Rn measurement system for the Super-K water.
Using this system, we measured the Rn concentration at several places in the Super-K tank and the water system 
with a background of $0.143\pm0.027~\mathrm{mBq/hour}$, which corresponds to $0.67\pm0.12~\mathrm{mBq/m^{3}}$ when $1~\mathrm{m^{3}}$ of water is sampled.
During the period from June 2014 to October 2015,
the Rn concentrations were stable at $1.74\pm0.14~\mathrm{mBq/m^{3}}$ in the supply water, 
$<0.23~\mathrm{mBq/m^{3}}$~($95\%$~C.L.) in the ID center, 
$2.63\pm0.22~\mathrm{mBq/m^{3}}$ at the bottom of the ID, 
and $9.06\pm0.58~\mathrm{mBq/m^{3}}$ in the return water.
Comparing the supply water with the return water, we conclude that the dominant Rn sources are in the Super-K tank 
rather than in the water system or buffer gas. 

The method developed in this study will enable other solar neutrino detectors, such as SK-Gd and Hyper-Kamiokande~\cite{hk}, 
to monitor the Rn concentrations in their purified water in detail.

\section*{Acknowledgements}
The authors would like to thank the Super-Kamiokande collaboration for their help in conducting this study. Especially, we thank M.~Miura who supported the scheduling to open the tank. We acknowledge the cooperation of the Kamioka Mining and Smelting Company. We would like to thank the reviewers for their thoughtful comments and careful review, which helped improve our manuscript. Y.~N thanks M.~Kanazawa for helping the work to assemble the measurement system at the experimental area of the Super-K. In addition, Y.~N thanks N.~Nozawa, T.~Onoue, Y.~Tamori, T.~Higashi and T.~Ushimaru for transporting air bottles. Y.~N thanks K.~Watanabe and H.~Nagao who supported the calibration works. Y.~N also thanks M.~Ikeda, T.~Yano, G.~Pronost, S.~Ito, K.~Iwamoto, Y.~Suda, A.~Orii, R.~Akutsu, D.~Fukuda, C.~Xu, K.~Ito and Y.~Sonoda who supported the water sampling work. This work is partially supported by the inter-university research program at ICRR. This work is partially supported by MEXT KAKENHI Grant Number 26104008, 17K17880, and 18H05536.

%\section*{References}


\begin{thebibliography}{00}

\bibitem{osci1} Ziro~Maki, Masami~Nakagawa, Shoichi~Sakata, Remarks on the Unified Model of Elementary Particles, Prog. Theor. Phys. 28 (1962) 870.

\bibitem{osci2} B.~Pontecorvo, Neutrino Experiments and the Problem of Conservation of Leptonic Charge, Soviet Physics JETP 26 (1968) 984--988.

\bibitem{sk1_es} S.~Fukuda, et al., Solar $\mathrm{^{8}B}$ and hep Neutrino Measurements from 1258 Days of Super-Kamiokande Data, Phys. Rev. Lett. 86 (2001) 5651. 

\bibitem{sno1} Q.R.~Ahmad, et al., Measurement of the Rate of $\nu_{e}+d\to p+p+e^{-}$ Interactions Produced by $\mathrm{^{8}B}$ Solar Neutrinos at the Sudbury Neutrino Observatory, Phys. Rev. Lett. 87 (2001) 071301. 

\bibitem{sno2} Q.R.~Ahmad, et al., Direct Evidence for Neutrino Flavor Transformation from Neutral-Current Interactions in the Sudbury Neutrino Observatory, Phys. Rev. Lett. 89 (2002) 011301. 


\bibitem{sk1} J.~Hosaka, et al., Solar neutrino measurements in Super-Kamiokande-I, Phys. Rev. D 73 (2006) 112001. 

\bibitem{sk2} J.P.~Cravens, et al., Solar neutrino measurements in Super-Kamiokande-II, Phys. Rev. D 78 (2008) 032002. 

\bibitem{sk3} K.~Abe, et al., Solar neutrino results in Super-Kamiokande-III, Phys. Rev. D 83 (2011) 052010. 


\bibitem{sk4_solar} K.~Abe, et al., Solar neutrino measurements in Super-Kamiokande-IV, Phys. Rev. D 94 (2016) 052010. 

\bibitem{msw1} S.P.~Mikheyev and A.Y.~Smirnov, Resonance enhancement of oscillations in matter and solar neutrino spectroscopy, Sov. J. Nucl. Phys. 42 (1985) 913--917.

\bibitem{msw2} L.~Wolfenstein, Neutrino oscillations in matter, Phys. Rev. D 17 (1978) 2369. 

\bibitem{dn_sk4} A.~Renshaw, et al., First Indication of Terrestrial Matter Effects on Solar Neutrino Oscillation, Phys. Rev. Lett. 112 (2014) 091805. 

\bibitem{Rn_sk} Y.~Takeuchi, et al., Measurement of radon concentrations at Super-Kamiokande, Phys. Lett. B 452 (1999) 418--424. 


\bibitem{paper_SK} S.~Fukuda, et al., The Super-Kamiokande detector, Nucl. Instrum. Meth. Phys. Res. Sect. A 501 (2003) 418--462. 

\bibitem{yamada} S.~Yamada, et al,. Commissioning of the new electronics and online system for the Super-Kamiokande experiment, IEEE Trans. Nucl. Sci. 57 (2010) 428--432.

\bibitem{Calib_p} K.~Abe, et al., Calibration of the Super-Kamiokande, Nucl. Instrum. Meth. Phys. Res. Sect. A 737 (2014) 253--272.

\bibitem{sno} B.~Aharmin, et al., Combined analysis of all three phases of solar neutrino data from the Sudbury Neutrino Observatory, Phys. Rev. C 88 (2013) 025501.

\bibitem{home} B.T.~Cleveland, et al., Measurement of the Solar Electron Neutrino Flux with the Homestake Chlorine Detector, Astrophys. J. 496 (1998) 505--526. 

\bibitem{sage} J.N.~Abdurashitov, et al., Measurement of the solar neutrino capture rate with gallium metal. III. Results for the 2002-2007 data-taking period, Phys. Rev. C 80 (2009) 015807.

\bibitem{gno} M.~Altmann, et al., Complete results for five years of GNO solar neutrino observations, Phys. Lett. B 616 (2005) 174--190. 

\bibitem{kamioka1} K.S.~Hirata, et al., Observation of $\mathrm{^{8}B}$ solar neutrinos in the Kamiokande-II detector, Phys. Rev. Lett. 63 (1989) 16.

\bibitem{kamioka2} Y.~Fukuda, et al., Solar Neutrino Data Covering Solar Cycle 22, Phys. Rev. Lett. 77 (1996) 1683. 

\bibitem{sno3} B.~Aharmim, et al., Electron energy spectra, fluxes, and day-night asymmetries of $\mathrm{^{8}B}$ solar neutrinos from measurements with NaCl dissolved in the heavy-water detector at the Sudbury Neutrino Observatory, Phys. Rev. C 72 (2005) 055502. 

\bibitem{sno4} B.~Aharmim, et al., Measurement of the $\nu_{e}$ and total $\mathrm{^{8}B}$ solar neutrino fluxes with the Sudbury Neutrino Observatory phase-III data set, Phys. Rev. C 87 (2013) 015502. 


\bibitem{bore1} G.~Bellini, et al., Measurement of the solar $\mathrm{^{8}B}$ neutrino rate with a liquid scintillator target and 3 MeV energy threshold in the Borexino detector, Phys. Rev. D 82 (2010) 033006. 

\bibitem{bore2} G.~Bellini, et al., First Evidence of $pep$ Solar Neutrinos by Direct Detection in Borexino, Phys. Rev. Lett. 108 (2012) 051302.

\bibitem{bore3} G.~Bellini, et al., Final results of Borexino Phase-I on low-energy solar neutrino spectroscopy, Phys. Rev. D 89 (2014) 112007.


\bibitem{bore4} G.~Bellini, et al., Neutrinos from the primary proton-proton fusion process in the Sun, Nature 512 (2014) 383--386.

\bibitem{bore5} M.~Agostini, et al., Simultaneous precision spectroscopy of $pp$, $^{7}$Be, and $pep$ solar neutrinos with Borexino Phase-II, Phys. Rev. D 100 (2019) 082004.

\bibitem{kamland2} S.~Abe, et al., Measurement of the $\mathrm{^{8}B}$ solar neutrino flux with the KamLAND liquid scintillator detector, Phys. Rev. C 84 (2011) 035804. 

\bibitem{kamland3} A.~Gando, et al., $\mathrm{^{7}Be}$ solar neutrino measurement with KamLAND, Phys. Rev. C 92 (2015) 055808. 

\bibitem{sterile1} P.C.~de~Holanda and A.Yu.~Smirnov, Homestake result, sterile neutrinos, and low energy solar neutrino experiments, Phys. Rev. D 69 (2004) 113002.

\bibitem{sterile2} P.C.~de~Holanda and A.Yu.~Smirnov, Solar neutrino spectrum, sterile neutrinos, and additional radiation in the Universe, Phys. Rev. D 83 (2011) 113011. 

\bibitem{sterile3} Ilidio Lopes, The Sterile-Active Neutrino Flavor Model: The Imprint of Dark Matter on the Electron Neutrino Spectra, Astrophys. J. 869 (2018) 112.

\bibitem{massvary} V.~Barger, Patrick~Huber, and Danny~Marfatia, Solar Mass-Varying Neutrino Oscillations, Phys. Rev. Lett. 95 (2005) 211802. 

\bibitem{non1} A.~Friedland, et al., Solar neutrinos as probes of neutrino-matter interactions, Phys. Lett. B 594 (2004) 347--354. 

\bibitem{non2} O.G.~Miranda, et al., Are solar neutrino oscillations robust?, J. High Energy Phys., 10 (2006) 008. 

\bibitem{Radon_takeuchi_1999} Y.~Takeuchi, et al., Development of high sensitivity radon detectors, Nucl. Instrum. Meth. Phys. Res. Sect. A 421 (1999) 334--341. 


\bibitem{sekiya} H.~Sekiya, Quest for the lowest-energy neutrinos in Super-Kamiokande, AIP Conf. Proc. 1672 (2015) 080001. 

\bibitem{sno_rn} I.~Blevis, et al., Measurement of $\mathrm{^{222}Rn}$ dissolved in water at the Subdury Neutrino Observatory, Nucl. Instrum. Meth. Phys. Res. Sect. A 517 (2003) 139--153. 

\bibitem{sno_ra1} T.C.~Andersen, et al., Measurement of radium concentration in water with Mn-coated beads at the Sudbury Neutrino Observatory, Nucl. Instrum. Meth. Phys. Res. Sect. A 501 (2003) 399--417. 

\bibitem{sno_ra2} B.~Aharmin, et al., High sensitivity measurement of $\mathrm{^{224}Ra}$ and $\mathrm{^{226}Ra}$ in water with an improved hydrous titanium oxide technique at the Sudbury Neutrino Observatory, Nucl. Instrum. Meth. Phys. Res. Sect. A 604 (2009) 531--535. 

\bibitem{bore_rn1} M.~Balata, et al., The water purification system for the low background counting test
facility of the Borexino experiment at Gran Sasso, Nucl. Instrum. Meth. Phys. Res. Sect. A 370 (1996) 605--608. 

\bibitem{bore_rn2} H.~Simgen, et al., A new system for the $^{222}\mathrm{Rn}$ and $^{226}\mathrm{Ra}$ assay of water and results in the Borexino project, Nucl. Instrum. Meth. Phys. Res. Sect. A 497 (2003) 407--413. 

\bibitem{dayabay} M.C.~Chu, et al., The radon monitoring system in Daya Bay Reactor Neutrino Experiment, Nucl. Instrum. Meth. Phys. Res. Sect. A 808 (2016) 156--164.

\bibitem{juno_rn} Y.P.~Zhang, et al., The development of $\mathrm{^{222}Rn}$ detectors for JUNO prototype, Radiat. Detect. Technol. Methods 2 (2018) 5. 

\bibitem{juno_rn2} L.~Xie, et al., Developing the radium measurement system for JUNO's water Cherenkov detector, arXiv:1906.06895.

\bibitem{Radon_takeuchi_2003} C.~Mitsuda, et al., Development of super-high sensitivity radon detector for the Super-Kamiokande detector, Nucl. Instrum. Meth. Phys. Res. Sect. A 497 (2003) 414--428. 

\bibitem{elect_method} P.~Kotrappa, S.K.~Dua, P.C.~Gupta, Y.S.~Mayya, Electret - A New Tool for Measuring Concentrations of Radon and Thoron in Air, Health Phys. 46 (1981) 35.

\bibitem{ynakano} Y.~Nakano, et al., Measurement of radon concentration in Super-Kamiokande's buffer gas, Nucl. Instrum. Meth. Phys. Res. Sect. A 867 (2017) 108--114. 

\bibitem{PTEP} K.~Hosokawa, et al., Development of a high-sensitivity 80 L radon detector for purified gases, Prog. Theor. Exp. Phys. 033H01 (2015). 

\bibitem{lsc3} J.~Steyn, Absolute Standardization of Beta-emitting Isotopes with a Liquid Scintillation Counter, Proc. Phys. Soci. Sec. A 69 (1956) 865. 

\bibitem{lsc2} M.~Noguchi, Special Applications (2), Measurements of Radon Activity, RADIOISOTOPES 24 (1975) 745--748.

\bibitem{lsc} Yumi~Yasuoka, et al., Determination of Radon Concentration in Water Using Liquid Scintillation Counter, RADIOISOTOPES 53 (2004) 123--131.

\bibitem{maurer} S.~Maurer, A~Mersmann and W.~Peukert, Henry coefficients of adsorption predicted from solid Hamaker constants, Chem. Eng. Sci. 56 (2001) 3443--3453. 

\bibitem{hardy} Hardy~Simgen, Adsorption techniques for gas purification, AIP Conf. Proc. 785 (2005) 121. 

\bibitem{xmass} K.~Abe, et al., Radon removal from gaseous xenon with activated charcoal, Nucl. Instrum. Meth. Phys. Res. Sect. A 661 (2012) 50--57.

\bibitem{lux} D.S.~Akerlib, et al., Chromatographic separation of radioactive noble gases from xenon, Astropart. Phys. 97 (2018) 80--87. 

\bibitem{motoyasu} M.~Ikeda, et al., Absorption and desorption of radon in argon gas, and the development of low level radon concentration measurement method, RADIOISOTOPES, 59 (2010) 29--36. 

\bibitem{charcoal} M.~Shimo, et al., Experimental Study of Charcoal Adsorptive Technique for Measurement of Radon in Air, J. Atom. Energy Soci. Jap. Vol. 25 (1983) 562--570.

\bibitem{humid} T.~Iida, et al.,An Electrostatic Integrating 222Rn Monitor with Cellulose Nitrate Film for Environmental Monitoring, Health Phys. 54 (1988) 139.

\bibitem{DT} E.~Blaufuss, et al., $\mathrm{^{16}N}$ as a calibration source for Super-Kamiokande, Nucl. Instrum. Meth. Phys. Res. Sect. A 458 (2001) 638--649. 

\bibitem{dron} Y.~Nakano, $^{8}\mathrm{B}$ solar neutrino spectrum measurement using Super-Kamiokande IV, PhD thesis, University of Tokyo (2016), available at http://www-sk.icrr.u-tokyo.ac.jp/sk/publications/index-e.html (accessed on October 9th, 2019).

\bibitem{nue2017} Y.~Nakano for the Super-Kamiokande Collaboration, Radon background study in Super-Kamiokande, J. Phys. Conf. Ser. 888 (2017) 012191. 


\bibitem{hk} K.~Abe, et al., Hyper-Kamiokande Design Report, arXiv:1805.04163.

%\end{linenumbers}
\end{thebibliography}
\end{document}